\documentstyle[fleqn,twoside,gc]{article}

\def\g{{\,\rm g}}
\def\eV{\,{\rm eV}}
\def\keV{\,{\rm keV}}
\def\MeV{\,{\rm MeV}}
\def\GeV{\,{\rm GeV}}

\def\sv{\left<\sigma v\right>}
\def\({\left(}
\def\){\right)}
\def\cm{{\,\rm cm}}

\heads{Belotsky, Khlopov, Shibaev}
      {Composite dark matter}

\begin{document}

\Title{COMPOSITE DARK MATTER  
       AND ITS CHARGED CONSTITUENTS}


   \Author{K.M. Belotsky\foom 1
   M.Yu. Khlopov\foom 2          
             and K.I.Shibaev\foom 3}   
          {Moscow Engineering Physics Institute, 115409 Moscow, Russia}             

\Abstract
    {Stable charged heavy leptons and quarks can exist and hide in elusive atoms, bound by Coulomb attraction
    and playing the role of dark matter. However, in the expanding
    Universe it is not possible to recombine all the charged particles into such atoms,
    and the positively charged particles, which
    escape this recombination, bind with electrons in atoms of
    anomalous isotopes with pregalactic abundance, exceeding
    substantially the terrestrial upper limits. This abundance can
    not be reduced in the dense matter bodies, if negatively
    charged particles have charge $-1$. Therefore
    composite dark matter can involve only negatively charged
    particles with charge $-2$, while stable heavy particles with charge
    $-1$ should be excluded.
    Realistic scenarios of composite dark matter, avoiding this problem of anomalous isotope
    over-production,
    inevitably predict the existence of primordial "atoms", in which primordial helium traps
    all the free negatively charged heavy constituents with charge $-2$. Study of the possibility for such primordial heavy $\alpha$ particle with
    compensated charge to exist as well as the search for the stable charged constituents in cosmic rays and accelerators provide
    crucial test for composite dark matter scenarios.}

\email 1 {k-belotsky@yandex.ru} \email 2
{maxim.khlopov@roma1.infn.it} \email 3 {shibaev01@yandex.ru}

\section{Introduction}

The problem of existence of new quarks and leptons is among the
most important in the modern high energy physics. This problem has
an interesting cosmological aspect. If these quarks and/or charged
leptons are stable, they should be present around us and the
reason for their evanescent nature should be found.

Recently at least three elementary particle frames for heavy
stable charged quarks and leptons were considered: (a) A heavy
quark and heavy neutral lepton (neutrino with mass above half the
Z-Boson mass) of fourth generation \cite{N,Legonkov}; which can
avoid experimental constraints \cite{Q,Okun} and form composite
dark matter species \cite{I,lom}; (b) A Glashow's ``Sinister''
heavy tera-quark $U$ and tera-electron $E$, which can form a tower
of tera-hadronic and tera-atomic bound states with ``tera-helium
atoms'' $(UUUEE)$ considered as dominant dark matter
\cite{Glashow,Fargion:2005xz}. Finally, (c) AC-leptons, predicted
in the extension \cite{5} of standard model, based on the approach
of almost-commutative geometry, \cite{book} can form evanescent
AC-atoms, playing the role of dark matter
\cite{5,FKS,Khlopov:2006uv}.

In all these recent models, predicting stable charged particles,
the particles escape experimental discovery, because they are
hidden in elusive atoms, maintaining dark matter of the modern
Universe. It offers new solution for the physical nature of the
cosmological dark matter.

This approach differs from the idea of dark matter, composed of
primordial bound systems of superheavy charged particles and
antiparticles, proposed earlier to explain the origin of Ultra
High Energy Cosmic Rays (UHECR) \cite{UHECR}. To survive to the
present time and to be simultaneously the source of UHECR
superheavy particles should satisfy a set of constraints, which in
particular exclude the possibility that they possess gauge charges
of the standard model.

The particles, considered here, participate in the standard model
interactions and we discuss the problems, related with various
dark matter scenarios with composite atom-like systems, formed by
heavy electrically charged stable particles.
\section{Grave shadows over the Sinister Universe}
In Glashow's  "Sinister" $SU(3)_c \times SU(2) \times SU(2)'
\times U(1)$ gauge model \cite{Glashow} three Heavy generations of
tera-fermions are related with the light fermions by $CP'$
transformation linking light fermions to charge conjugates of
their heavy partners and vice versa. $CP'$ symmetry breaking makes
tera-fermions much heavier than their light partners. Tera-fermion
mass pattern is the same as for light generations, but all the
masses are multiplied by the same factor $S =10^6 S_6 \sim 10^6$.
Strict conservation of $F = (B-L) - (B' - L')$ prevents mixing of
charged tera-fermions with light quarks and leptons. Tera-fermions
are sterile relative to $SU(2)$ electroweak interaction, and do
not contribute into standard model parameters.
 In such realization the new heavy neutrinos ($N_i$)
  acquire large masses and their
mixing with light neutrinos $\nu$ provides a "see-saw" mechanism
of light neutrino Dirac mass generation. Here in a Sinister model
the heavy neutrino is unstable. On the contrary in this scheme
$E^-$ is the lightest heavy fermion and it is absolutely stable.

Since the lightest quark $U$ of Heavy generation does not mix with
quarks of 3 light generations, it can decay only to Heavy
generation leptons owing to GUT-type interactions, what makes it
sufficiently long living. If its lifetime exceeds the age of the
Universe, primordial $U$-quark hadrons as well as Heavy Leptons
$E^-$ 
should be present in the modern matter.

Glashow's "Sinister" scenario \cite{Glashow} took into account
that
 very heavy quarks $Q$ (or antiquarks $\bar Q$) can form bound states with other heavy quarks
 (or antiquarks) due to their Coulomb-like QCD attraction, and the binding energy of these states
 substantially exceeds the binding energy of QCD confinement.
Then $(QQq)$ and $(QQQ)$ baryons can exist.

According to  \cite{Glashow} primordial heavy quark $U$ and heavy
electron $E$ are stable and
may form a neutral most probable and stable (while being
evanescent) $(UUUEE)$ "atom"
with $(UUU)$ hadron as nucleus and two $E^-$s as "electrons". The
tera gas of such "atoms" seemed an ideal candidate for a very new
and fascinating dark matter;
because of their peculiar WIMP-like interaction with matter they
might also rule the stages of gravitational clustering in early
matter dominated epochs, creating first gravity seeds for galaxy
formation.

The problem of such scenario is inevitable presence of "products
of incomplete combustion" and the necessity to decrease their
abundance. Indeed in analogy to D, $^3$He and Li relics that are
the intermediate catalyzers of $^4$He formation in Standard Big
Bang Nucleosynthesis (SBBN) and are important cosmological tracers
of this process, the tera-lepton and tera-hadron relics from
intermediate stages of a multi-step process of towards a final
$(UUUEE)$ formation must survive with high abundance of {\it
visible} relics in the present Universe. To avoid this trouble an
original idea of $(Ep)$ catalysis was proposed in \cite{Glashow}:
as soon as the temperature falls down below $T\sim I_{Ep}/25 \sim
1 \keV$ neutral $(Ep)$ atom with "ionization potential"
$I_{Ep}=\alpha^2m_p/2=25 \keV$ can be formed. The hope was
\cite{Glashow} that this "atom" must catalyze additional effective
binding of various tera-particle species and to reduce their
abundance below the experimental upper limits.

Unfortunately, as it was shown in \cite{Fargion:2005xz}, this
fascinating picture of Sinister Universe can not be realized.
Tracing in more details cosmological evolution of tera-matter and
strictly following the conjecture of \cite{Glashow}, the troubles
of this approach were revealed and gracious exit from them for any
model assuming -1 charge component of composite atom-like dark
matter was found impossible.

The model \cite{Glashow} didn't offer any physical mechanism for
generation of cosmological tera-baryon asymmetry and such
asymmetry was postulated to saturate the observed dark matter
density. This assumption was taken in \cite{Fargion:2005xz} and it
was revealed that while the assumed tera-baryon asymmetry for $U$
washes out by annihilation primordial $\bar U$, the tera-lepton
asymmetry of $E^-$ can not effectively suppress the abundance of
tera-positrons $E^+$ in the earliest as well as in the late
Universe stages. This feature differs from successful annihilation
of primordial antiprotons and positrons that takes place in our
Standard baryon asymmetrical Universe. The abundance  of $\bar U$
and $E^+$ in earliest epochs exceeds the abundance of excessive
$U$ and $E^-$ and it is suppressed (successfully) for $\bar U$
only after QCD phase transition, while, there is no such effective
annihilation mechanism for $E^+$. Thus the tera-lepton pair
overproduction was revealed as the first trouble of Sinister
Universe.

Moreover ordinary $^4$He formed in Standard Big Bang
Nucleosynthesis binds at $T \sim 15 keV$ virtually all the free
$E^-$ into positively charged $(^4HeE^-)^+$ "ion", which puts
Coulomb barrier for any successive $E^-E^+$ annihilation or any
effective $EU$ binding. It happens {\it before} $(Ep)$ atom can be
formed and $(Ep)$ atoms can not be formed, since all the free $E$
are already imprisoned by $^4$He cage. It removed the hope
\cite{Glashow} on $(Ep)$ atomic catalysis as {\it panacea} from
unwanted tera-particle species and became the second unresolvable
trouble for the Sinister Universe.

 The huge frozen abundance of tera-leptons in hybrid tera-positronium $(eE^+)$
 and hybrid hydrogen-like tera-helium atom $(^4He Ee)$ and
 in other complex anomalous isotopes can not be removed \cite{Fargion:2005xz}.

Their abundance is enormously high for known severe bounds on
anomalous hydrogen. This is the grave nature of tera-lepton
shadows over a Sinister Universe.

The remaining abundance of $(eE^+)$ and $(^4HeE^-e)$ exceeds by
{\it 27 orders} of magnitude the terrestrial upper limit for
anomalous hydrogen. There are also additional tera-hadronic
anomalous relics, whose trace is constrained by the present data
by  {\it 25.5 orders} for $(UUUEe)$ and {\it at least by 20
orders} for $(Uude)$ respect to anomalous hydrogen ($r<10^{-30}$
relative to atom number density in Earth), as well as by {\it 14.5
orders} for $(UUUee)$, by {\it 10 orders} for $(UUuee)$ - respect
to anomalous helium ($r<10^{-19}$). While tera helium $(UUUEE)$
would co-exist with observational data, being a wonderful
candidate for dark matter, its tera-lepton partners poison and
forbid this opportunity.

The contradiction might be removed, if tera-fermions are unstable
and drastically decay before the present time. But such solution
excludes any cosmological sinister matter dominated Universe,
while, of course, it leaves still room and challenge for search
for metastable $E$-leptons and $U$-hadrons in laboratories or in
High Energy Cosmic ray traces.
\section{Composite dark matter from almost commutative geometry}
The AC-model \cite{5} appeared as realistic elementary particle
model, based on the specific approach of \cite{book} to unify
general relativity, quantum mechanics and gauge symmetry.

This realization naturally embeds the Standard model, both
reproducing its gauge symmetry and Higgs mechanism, but to be
realistic, it should go beyond the standard model and offer
candidates for dark matter. Postulates of noncommutative geometry
put severe constraints on the gauge symmetry group, excluding in
this approach, which can be considered as alternative to
superstring phenomenology, supersymmetric and GUT extensions. The
AC-model \cite{5} extends the fermion content of the Standard
model by two heavy particles with opposite electromagnetic and
Z-boson charges. Having no other gauge charges of Standard model,
these particles (AC-fermions) behave as heavy stable leptons with
charges $-2e$ and $+2e$, called here $A$ and $C$, respectively.
AC-fermions are sterile relative to $SU(2)$ electro-weak
interaction, and do not contribute to the standard model
parameters. The mass of AC-fermions is originated from
noncommutative geometry of the internal space (thus being much
less than the Planck scale) and is not related to the Higgs
mechanism. The lower limit for this mass follows from absence of
new chrged leptons in LEP. It was assumed in
\cite{FKS,Khlopov:2006uv} that $m_A = m_C = m = 100 S_2{\GeV} $
with free parameter $S_2 \ge 1$.In the absence of AC-fermion
mixing with light fermions, AC-fermions can be absolutely stable.
Such absolute stability and absence of mixing with ordinary
particles naturally follows from strict conservation of additional
$U(1)$ gauge charge, which is called $y$-charge and which only
AC-leptons possess \cite{FKS,Khlopov:2006uv}.

If AC-leptons $A$ and $C$ have equal and opposite sign
$y$-charges, strict conservation of $y$-charge does not prevent
generation of $A$ and $C$ excess, the excess of $A$ being equal to
excess of $C$. The mechanism of baryosynthesis in the present
version of AC model is not clear, therefore the AC-lepton excess
was postulated in \cite{5,FKS,Khlopov:2006uv} to saturate the
modern CDM density (similar to the approach sinister model).
 Primordial excessive negatively charged $A^{--}$ and positively
charged $C^{++}$ form a neutral most probable and stable (while
being evanescent) $(AC)$ "atom", the AC-gas of such "atoms" being
a candidate for dark matter \cite{5,FKS,Khlopov:2006uv}.

However, similar to the sinister Universe AC-lepton relics from
intermediate stages of a multi-step process towards a final $(AC)$
atom formation must survive with high abundance of {\it visible}
relics in the present Universe. In spite of the assumed excess of
particles ($A^{--}$ and $C^{++}$) abundance of frozen out
antiparticles ($\bar A^{++}$ and $\bar C^{--}$) is not negligible,
as well as significant fraction of $A^{--}$ and $C^{++}$ remains
unbound, when $AC$ recombination takes place and most of
AC-leptons form $(AC)$ atoms. As soon as $^4He$ is formed in Big
Bang nucleosynthesis it captures all the free negatively charged
heavy particles. If the charge of such particles is -1 (as it was
the case for tera-electron in \cite{Glashow}) positively charged
ion $(^4He^{++}E^{-})^+$ puts Coulomb barrier for any successive
decrease of abundance of species, over-polluting by anomalous
isotopes modern Universe. This problem of unavoidable
over-abundance of by-products of "incomplete combustion" is
avoided in AC-model owing to the double negative charge of
$A^{--}$ \cite{FKS,Khlopov:2006uv}. Instead of positively charged
ion the primordial component of free anion-like AC-leptons
$A^{--}$ are mostly trapped in the first three minutes into a
puzzling neutral OLe-helium state (named so from \emph{O-Le}pton-
\emph{helium}) $(^4He^{++}A^{--})$, with nuclear interaction cross
section, which provides anywhere eventual later $(AC)$ binding. As
soon as OLe-helium forms, it catalyzes in first three minutes
effective binding in $(AC)$ atoms and complete annihilation of
antiparticles. Products of annihilation cause undesirable effect
neither in CMB spectrum, nor in light element abundances. Due to
early decoupling from relativistic plasma y-photon background is
suppressed and its contribution to the total density in the period
of Big Bang Nucleosynthesis is compatible with observational
constraints. OLe-helium, this surprising catalyzer with screened
Coulomb barrier, can influence the chemical evolution of ordinary
matter, but as we argue further the dominant process of OLe-helium
interaction with nuclei might be quasi-elastic and might not
result in over-production of anomalous isotopes.

The development of gravitational instabilities of AC-atomic gas
follows the general path of the CDM scenario, but the composite
nature of $(AC)$-atoms leads to some specific difference. For
$S_2<6$ the bulk of $(AC)$ bound states appear in the Universe at
$T_{fAC} = 0.7 S_2 \MeV$ and the minimal mass of their
gravitationally bound systems is given by the total mass of $(AC)$
within the cosmological horizon in this period, which is of the
order of $M = (T_{RM}/T_{fAC}) m_{Pl} (m_{Pl}/T_{fAC})^2 \approx
6\cdot 10^{33}/S_2^3 \g, $ where $T_{RM}=1 \eV$ corresponds to the
beginning of the AC-matter dominated stage. At $S_2>6$ the bulk of
$(AC)$-atoms is formed only at $T_{OHe} = 60 \keV$ due to
OLe-helium catalysis. Therefore at $S_2>6$ the minimal mass is
independent of $S_2$ and is given by $M = (T_{RM}/T_{OHe}) m_{Pl}
(m_{Pl}/T_{OHe})^2 \approx 10^{37} \g. $

At small energy transfer $\Delta E \ll m$ cross section for
interaction of AC-atoms with matter is suppressed by the factor
$\sim Z^2 (\Delta E/m)^2$, being for scattering on nuclei with
charge $Z$ and atomic weight $A$ of the order of $\sigma_{ACZ}
\sim Z^2/\pi (\Delta E/m)^2 \sigma_{AC} \sim Z^2 A^2 10^{-43}
\cm^2 /S^2_2.$ Here we take $\Delta E \sim 2 A m_p v^2$ and $v/c
\sim 10^{-3}$ and find that even for heavy nuclei with $Z \sim
100$ and $A \sim 200$ this cross section does not exceed $4 \cdot
10^{-35} \cm^2 /S^2_2.$ It proves WIMP-like behavior of AC-atoms
in the ordinary matter. In the Galaxy they behave as collisionless
gas.

Still, though CDM in the form of $(AC)$ atoms is successfully
formed, $A^{--}$ (bound in OLe-helium) and $C^{++}$ (forming
anomalous helium atom $(eeC^{++})$) should be also present in the
modern Universe and the abundance of primordial $(eeC^{++})$ is by
up to {\it ten} orders of magnitude higher, than experimental
upper limit on the anomalous helium abundance in terrestrial
matter. This problem can be solved by OLe-helium catalyzed $(AC)$
binding of $(eeC^{++})$, but different mobilities in matter of
atomic interacting $(eeC^{++})$ and nuclear interacting $(OHe)$
lead to fractionating of these species, preventing effective
decrease of anomalous helium abundance. The $U(1)$ charge
neutrality condition naturally prevents this fractionating, making
$(AC)$ binding sufficiently effective to suppress terrestrial
anomalous isotope abundance below the experimental upper limits.
Inside dense matter objects (stars or planets) its recombination
with $(eeC^{++})$ into $(AC)$ atoms can provide a mechanism for
the formation of dense $(AC)$ objects. In this process OLe-helium
and anomalous helium, which were coupled to the ordinary matter by
hadronic and atomic interactions, convert into $(AC)$ atoms, which
immediately sinks down to the center of the body.

However, though $(AC)$ binding is not accompanied by strong
annihilation effects, as it was the case for 4th generation
hadrons \cite{Q}, gamma radiation from it inside large volume
detectors should take place. In the course of $(AC)$ atom
formation electromagnetic transitions with $\Delta E > 1 \MeV$ can
be a source of $e^+e^-$ pairs, either directly with probability
$\sim 10^{-2}$ or due to development of electromagnetic cascade.
If $AC$ recombination goes on homogeneously in Earth within the
water-circulating surface layer of the depth $L \sim 4 \cdot 10^5
\cm$ inside the volume of Super Kamiokande with size $l_{K} \sim 3
\cdot 10^3 \cm$ equilibrium $AC$ recombination should result in a
flux of $e^+e^-$ pairs $F_e = N_e I_C l_{K}/L$, which for $N_e
\sim 1$ can be as large as $F_e \sim \cdot
\frac{10^{-12}}{f(S_2)}\frac{S_h}{5 \cdot 10^{-5}} (cm^2 \cdot s
\cdot ster)^{-1}.$ Their signal might be easily  disentangled
\cite{FKS,Khlopov:2006uv}(above a few MeV range) respect common
charged
    current neutrino interactions and single electron tracks
     because the tens MeV gamma lead, by pair productions, to twin
    electron  tracks,  nearly aligned along their Cerenkov rings.
    The signal is piling  the energy in windows where
   few atmospheric neutrino and  cosmic Super-Novae  radiate. The predicted signal
strongly depends, however, on the uncertain astrophysical
parameters \cite{FKS,Khlopov:2006uv}.

In this way AC-cosmology escapes most of the troubles, revealed
for other cosmological scenarios with stable heavy charged
particles \cite{Q,Fargion:2005xz} and provides realistic scenario
for composite dark matter in the form of evanescent atoms,
composed by heavy stable electrically charged particles, bearing
the source of invisible light.

\section{Stable pieces of 4th generation matter}
Precision data on Standard model parameters admit \cite{Okun} the
existence of 4th generation, if 4th neutrino ($N$) has mass about
50 GeV, while masses of other 4th generation particles are close
to their experimental lower limits, being $>100$GeV for charged
lepton ($E$) and $>300$GeV for 4th generation $U$ and $D$ quarks
\cite{CDF}. The results of this analysis determine our choice for
masses of $N$ ($m_N=50$ GeV) and $U$ ($m_U=350 S_5$ GeV).

4th generation can follow from heterotic string phenomenology and
its difference from the three known light generations can be
explained by a new gauge charge ($y$-charge), possessed only by
its quarks and leptons \cite{N,Q,I}. Similar to electromagnetism
this charge is the source of a long range Coulomb-like
$y$-interaction. Strict conservation of $y$-charge makes the
lightest particle of 4th family (4th neutrino $N$) absolutely
stable, while the lightest quark must be sufficiently long living
\cite{Q,I}. The lifetime of $U$ can exceed the age of the
Universe, as it was revealed in \cite{Q,I} for $m_U<m_D$.

The $y$-charges ($Q_y$) of ($N,E,U,D$) are fixed by the following
conditions. Cancellation of $Z-\gamma-y$ anomaly implies $Q_{yE} +
2\cdot Q_{yU}+ Q_{yD} = 0$; while cancellation of $Z-y-y$ anomaly
needs $Q_{yN}^2 - Q_{yE}^2 +3\cdot(Q_{yU}^2 - Q_{yD}^2) = 0.$
Proper $N-E$ and $U-D$ transitions of weak interaction assume
$Q_{yN} = Q_{yE}$ and $Q_{yU} = Q_{yD}$. From these conditions
follows $Q_{yE}=Q_{yN}=-3\cdot Q_{yU}=-3\cdot Q_{yD}$ so that
$y$-charges of ($N,E,U,D$) are ($1,1,-1/3,-1/3$).

$U$-quark can form lightest $(Uud)$ baryon and $(U \bar u)$ meson.
The corresponding antiparticles are formed by $\bar U$ with light
quarks and antiquarks. Owing to large chromo-Coulomb binding
energy ($\propto \alpha_{c}^2 \cdot m_U$, where $\alpha_{c}$ is
the QCD constant) stable double and triple $U$ bound states
$(UUq)$, $(UUU)$ and their antiparticles $(\bar U \bar U \bar u)$,
$(\bar U \bar U \bar U)$ can exist
\cite{Q,Glashow,Fargion:2005xz}. Formation of these double and
triple states in particle interactions at accelerators and in
cosmic rays is strongly suppressed, but they can form in early
Universe and strongly influence cosmological evolution of 4th
generation hadrons. As shown in \cite{I}, \underline{an}ti-
\underline{U}-\underline{t}riple state called \underline{anut}ium
or $\Delta^{--}_{3 \bar U}$ is of special interest. This stable
anti-$\Delta$-isobar, composed of $\bar U$ antiquarks and bound by
chromo-Coulomb force has the size $r_{\Delta} \sim
1/(\alpha_{QCD}\cdot m_U)$, which is much less than normal
hadronic size $r_{h} \sim 1/m_{\pi}$.

The model  \cite{Q} admits that in the early Universe an
antibaryon asymmetry for 4th generation quarks can be generated
\cite{I,lom}. Due to $y$-charge conservation $\bar U$ excess
should be compensated by $\bar N$ excess. $\bar U$-antibaryon
density can be expressed through the modern dark matter density
$\Omega_{\bar U}= k \cdot \Omega_{CDM}=0.224$ ($k \le 1$),
saturating it at $k=1$. It is convenient
\cite{Glashow,Fargion:2005xz,FKS,I,Khlopov:2006uv,lom} to relate
the baryon (corresponding to $\Omega_b=0.044$) and $\bar U$ ($\bar
N$) excess with the entropy density $s$, introducing $r_b = n_b/s$
and $r_{\bar U}=n_{\bar U}/s=3 \cdot n_{\bar N}/s=3 \cdot r_{\bar
N}$. One obtains $r_b \sim 8 \cdot 10^{-11}$ and $r_{\bar U},$
corresponding to $\bar U$ excess in the early Universe
$\kappa_{\bar U} =r_{\bar U} -r_{U}= 3 \cdot (r_{\bar N}
-r_{N})=10^{-12} (350 \GeV/m_U) = 10^{-12}/S_5,$ where $S_5 =
m_U/350{\GeV}$.
\subsection{Primordial composite forms of 4th generation matter}\label{subsec:composite}
In the early Universe at temperatures highly above their masses
$\bar U$ and $\bar N$ were in thermodynamical equilibrium with
relativistic plasma. It means that at $T>m_U$ ($T>m_N$) the
excessive $\bar U$ ($\bar N$) were accompanied by $U \bar U$ ($N
\bar N$) pairs.

Due to $\bar U$ excess frozen out concentration of deficit
$U$-quarks is suppressed at $T<m_U$ for $k>0.04$ \cite{lom}. It
decreases further exponentially first at $T \sim I_U \approx \bar
\alpha^2 M_U/2 \sim 3  S_5$GeV (where \cite{Q} $\bar \alpha= C_F
\alpha_{c} = 4/3 \cdot 0.144 \approx 0.19$ and $M_U = m_U/2$ is
the reduced mass), when the frozen out $U$ quarks begin to bind
with antiquarks $\bar U $ into charmonium-like state $(\bar U U)$
and annihilate. On this line $\bar U$ excess binds at $T < I_U$ by
chromo-Coulomb forces dominantly into $(\bar U \bar U \bar U)$
anutium states with electric charge $Z_{\Delta}=-2$ and mass
$m_o=1.05 S_5$TeV, while remaining free $\bar U$ anti-quarks and
anti-diquarks $(\bar U \bar U)$ form after QCD phase transition
normal size hadrons $(\bar U u)$ and $(\bar U \bar U \bar u)$.
Then at $T = T_{QCD} \approx 150$MeV additional suppression of
remaining $U$-quark hadrons takes place in their hadronic
collisions with $\bar U$-hadrons, in which $(\bar U U)$ states are
formed and $U$-quarks successively annihilate.

Owing to weaker interaction effect of $\bar N$ excess in the
suppression of deficit N is less pronounced and it takes place at
$T<m_N$ only for $k>0.002$ \cite{lom}. At $T \sim I_{NN} =
\alpha_y^2 M_N/4 \sim 15 $MeV (for $\alpha_y = 1/30$ and
$M_N=50$GeV) due to $y$-interaction the frozen out $N$ begin to
bind with $\bar N $ into charmonium-like states $(\bar N N)$ and
annihilate. At $T < I_{NU} = \alpha_y^2 M_N/2 \sim 30 $MeV
$y$-interaction causes binding of $N$ with $\bar U$-hadrons
(dominantly with anutium) but only at $T \sim I_{NU}/30 \ 1$MeV
this binding is not prevented by back reaction of
$y$-photo-destruction.

To the period of Standard Big Bang Nucleosynthesis (SBBN) $\bar U$
are dominantly bound in anutium $\Delta^{--}_{3 \bar U}$ with
small fraction ($\sim 10^{-6}$) of neutral $(\bar U u)$ and doubly
charged $(\bar U \bar U \bar u)$ hadron states. The dominant
fraction of anutium is bound by $y$-interaction with $\bar N$ in
$(\bar N \Delta^{--}_{3 \bar U})$ "atomic" state. Owing to early
decoupling of $y$-photons from relativistic plasma presence of
$y$-radiation background does not influence SBBN processes
\cite{Q,I}.

At $T<I_{o} = Z^2 Z_{He}^2 \alpha^2 m_{He}/2 \approx 1.6$MeV the
reaction $\Delta^{--}_{3 \bar U}+^4He\rightarrow \gamma
+(^4He^{++}\Delta^{--}_{3 \bar U})$ might take place, but it can
go only after $^4He$ is formed in SBBN at $T<100 $keV and is
effective only at $T \le T_{rHe} \sim
I_{o}/\log{\left(n_{\gamma}/n_{He}\right)} \approx I_{o}/27
\approx 60 $keV, when the inverse reaction of photo-destruction
cannot prevent it \cite{Fargion:2005xz,FKS,I,Khlopov:2006uv}. In
this period anutium is dominantly bound with $\bar N$. Since
$r_{He}=0.1 r_{b} \gg r_{\Delta}= r_{\bar U}/3 $, in this reaction
all free negatively charged particles are bound with helium
\cite{Fargion:2005xz,FKS,I,Khlopov:2006uv} and neutral
Anti-Neutrino-O-helium (ANO-helium, $ANOHe$) $(^4He^{++} [\bar N
\Delta^{--}_{3 \bar U}])$ ``molecule'' is produced with mass
$m_{OHe} \approx m_o \approx 1S_5$TeV. The size of this
``molecule'' is $ R_{o} \sim 1/(Z_{\Delta} Z_{He}\alpha m_{He})
\approx 2 \cdot 10^{-13}$ cm
 and it can play the role of a dark matter component and
a nontrivial catalyzing role in nuclear transformations.

In nuclear processes ANO-helium looks like an $\alpha$ particle
with shielded electric charge. It can closely approach nuclei due
to the absence of a Coulomb barrier and opens the way to form
heavy nuclei in SBBN. This path of nuclear transformations
involves the fraction of baryons not exceeding $10^{-7}$ \cite{I}
and it can not be excluded by observations.

\subsection{ANO-helium catalyzed processes}\label{subsec:O-helium}

As soon as ANO-helium is formed, it catalyzes annihilation of
deficit $U$-hadrons and $N$. Charged $U$-hadrons penetrate neutral
ANO-helium, expel $^4He$, bind with anutium and annihilate falling
down the center of this bound system. The rate of this reaction is
$\sv= \pi R^2_o$ and an $\bar U$ excess $k=10^{-3}$ is sufficient
to reduce the primordial abundance of $(Uud)$ below the
experimental upper limits. $N$ capture rate is determined by the
size of $(\bar N \Delta)$ "atom" in ANO-helium and its
annihilation is less effective.

The size of ANO-helium is of the order of the size of $^4He$ and
for a nucleus A with electric charge $Z>2$ the size of the Bohr
orbit for a $(Z \Delta)$ ion is less than the size of nucleus A.
This means that while binding with a heavy nucleus $\Delta$
penetrates it and effectively interacts with a part of the nucleus
with a size less than the corresponding Bohr orbit. This size
corresponds to the size of $^4He$, making O-helium the most bound
$(Z \Delta)$-atomic state.

The cross section for $\Delta$ interaction with hadrons is
suppressed by factor $\sim (p_h/p_{\Delta})^2 \sim
(r_{\Delta}/r_h)^2 \approx 10^{-4}/S_5^2$, where $p_h$ and
$p_{\Delta}$ are quark transverse momenta in normal hadrons and in
anutium, respectively. Therefore anutium component of $(ANOHe)$
can hardly be captured and bound with nucleus due to strong
interaction. However, interaction of the $^4He$ component of
$(ANOHe)$ with a $^A_ZQ$ nucleus can lead to a nuclear
transformation due to the reaction $^A_ZQ+(\Delta He) \rightarrow
^{A+4}_{Z+2}Q +\Delta,$ provided that the masses of the initial
and final nuclei satisfy the energy condition $M(A,Z) + M(4,2) -
I_{o}> M(A+4,Z+2),$ where $I_{o} = 1.6$MeV is the binding energy
of O-helium and $M(4,2)$ is the mass of the $^4He$ nucleus. The
final nucleus is formed in the excited $[\alpha, M(A,Z)]$ state,
which can rapidly experience $\alpha$- decay, giving rise to
$(ANOHe)$ regeneration and to effective quasi-elastic process of
$(ANOHe)$-nucleus scattering. It leads to possible suppression of
ANO-helium catalysis of nuclear transformations in matter.
\subsection{ANO-helium dark matter}\label{subsec:matter}
At $T < T_{od} \approx 1 \keV$ energy and momentum transfer from
baryons to ANO-helium $n_b \sv (m_p/m_o) t < 1$ is not effective.
Here $\sigma \approx \sigma_{o} \sim \pi R_{o}^2 \approx
10^{-25}\cm^2.$ and $v = \sqrt{2T/m_p}$ is baryon thermal
velocity. Then ANO-helium gas decouples from plasma and radiation
and plays the role of dark matter, which starts to dominate in the
Universe at $T_{RM}=1 \eV$.

The composite nature of ANO-helium makes it more close to warm
dark matter. The total mass of $(OHe)$ within the cosmological
horizon in the period of decoupling is independent of $S_5$ and
given by $$M_{od} = \frac{T_{RM}}{T_{od}} m_{Pl}
(\frac{m_{Pl}}{T_{od}})^2 \approx 2 \cdot 10^{42} \g = 10^9
M_{\odot}. $$ O-helium is formed only at $T_{o} = 60 \keV$ and the
total mass of $OHe$ within cosmological horizon in the period of
its creation is $M_{o}=M_{od}(T_{o}/T_{od})^3 = 10^{37} \g$.
Though after decoupling Jeans mass in $(OHe)$ gas falls down $M_J
\sim 3 \cdot 10^{-14}M_{od}$ one should expect strong suppression
of fluctuations on scales $M<M_o$ as well as adiabatic damping of
sound waves in RD plasma for scales $M_o<M<M_{od}$. It provides
suppression of small scale structure in the considered model. This
dark matter plays dominant role in formation of large scale
structure at $k>1/2$.

The first evident consequence of the proposed scenario is the
inevitable presence of ANO-helium in terrestrial matter, which is
opaque for $(ANOHe)$ and stores all its in-falling flux. If its
interaction with matter is dominantly quasi-elastic, this flux
sinks down the center of Earth. If ANO-helium regeneration is not
effective and $\Delta$ remains bound with heavy nucleus $Z$,
anomalous isotope of $Z-2$ element appears. This is the serious
problem for the considered model.

Even at $k=1$ ANO-helium gives rise to less than 0.1 \cite{I,lom}
of expected background events in XQC experiment \cite{XQC}, thus
avoiding for all $k \le 1$ severe constraints on Strongly
Interacting Massive particles SIMPs obtained in
\cite{McGuire:2001qj} from the results of this experiment. In
underground detectors $(ANOHe)$ ``molecules'' are slowed down to
thermal energies far below the threshold for direct dark matter
detection. However, $(ANOHe)$ destruction can result in observable
effects. Therefore a special strategy in search for this form of
dark matter is needed. An interesting possibility offers
development of superfluid $^3He$ detector
\cite{Winkelmann:2005un}. Due to high sensitivity to energy
release above ($E_{th} = 1 \keV$), operation of its actual few
gram prototype can put severe constraints on a wide range of $k$
and $S_5$.

At $10^{-3}<k<0.02$ $U$-baryon abundance is strongly suppressed
\cite{lom}, while the modest suppression of primordial $N$
abundance does not exclude explanation of DAMA, HEAT and EGRET
data in the framework of hypothesis of 4th neutrinos \cite{N} but
makes the effect of $N$ annihilation in Earth consistent with the
experimental data.

\section{Discussion}
To conclude, the existence of heavy stable charged particles can
offer new solution for dark matter problem. Dark matter candidates
can be atom-like states, in which negatively and positively stable
charged particles are bound by Coulomb attraction. Primordial
excess of these particles over their antiparticles implies the
mechanism of its generation and is still an open problem for all
the considered models. However, even if such mechanism exists and
the generated particle excess can prevent undesirable effect of
their annihilation and provide hiding of most of stable charged
particles into neutral WIMP-like states, there is a serious
problem of accompanying anomalous forms of atomic matter.

Indeed, recombination of charged species is never complete in the
expanding Universe, and significant fraction of free charged
particles should remain unbound. Free positively charged species
behave as nuclei of anomalous isotopes, giving rise to a danger of
their over-production. Moreover, as soon as $^4He$ is formed in
Big Bang nucleosynthesis it captures all the free negatively
charged heavy particles. If the charge of such particles is -1 (as
it is the case for tera-electron in \cite{Glashow}) positively
charged ion $(^4He^{++}E^{-})^+$ puts Coulomb barrier for any
successive decrease of abundance of species, over-polluting modern
Universe by anomalous isotopes. It excludes the possibility of
composite dark matter with $-1$ charged constituents and only $-2$
charged constituents avoid these troubles, being trapped by helium
in neutral OLe-helium or O-helium (ANO-helium) states.

The existence of $-2$ charged states and the absence of stable
$-1$ charged constituents can take place in AC-model and in charge
asymmetric model of 4th generation. In the first case, pregalactic
abundance of $C^{++}$ exceeds by ten orders of magnitude the
terrestrial upper limit on anomalous helium and the mechanism of
suppression of this abundance is inevitably accompanied by
observable effects of recombination and implies the existence of
$y$ charge, possessed by AC leptons. In the second case, owing to
excess of $\bar U$ anti-quarks primordial abundance of positively
charged $U$-baryons is exponentially suppressed and anomalous
isotope over-production is avoided. Excessive anti-$U$-quarks
should retain dominantly in the form of anutium, which binds with
excessive $\bar N$ and then with $^4He$ in neutral ANO-helium. In
the both cases, OLe-helium (or ANO-helium) should exist and its
possible role in nuclear transformation is the serious danger (or
exciting advantage?) for composite dark matter scenario.

Galactic cosmic rays destroy ANO-helium (as well as OLe-helium),
striking off $^4He$. It can lead to appearance of a free
[anutium-$\bar N$] component in cosmic rays, which can be as large
as $[\bar N \Delta^{--}_{3 \bar U}]/^4He \sim 10^{-7}$ and
accessible to PAMELA and AMS experiments. The estimation is two
orders less in the case of free $A^{--}$ from cosmic ray
destruction of OLe-helium \cite{Khlopov:2006uv}.

In the context of composite dark matter like \cite{I,lom} or
\cite{FKS,Khlopov:2006uv} accelerator search for new stable quarks
and leptons acquires the meaning of critical test for existence of
its charged components. Such test will be possible in experiment
ATLAS/LHC (see Fig.1). The search can be realized with the use of
time-of-light technique. The quest particles can be discriminated
by low velocity $v/c < 1$ and high transverse momentum. Moreover,
in the case of hadronized heavy stable $U$-quarks
components of $U\bar{U}$-pair born in $pp$-collision will fly away
basically in form of positively charged $U$-baryon and
$\bar{U}$-meson flipping their charge during passage through
detectors ($U$-meson and $\bar{U}$-baryon will convert to
respectively $U$-baryon and $\bar{U}$-meson within a few nuclear
lengths from the bunch crossing point).

\EFigure{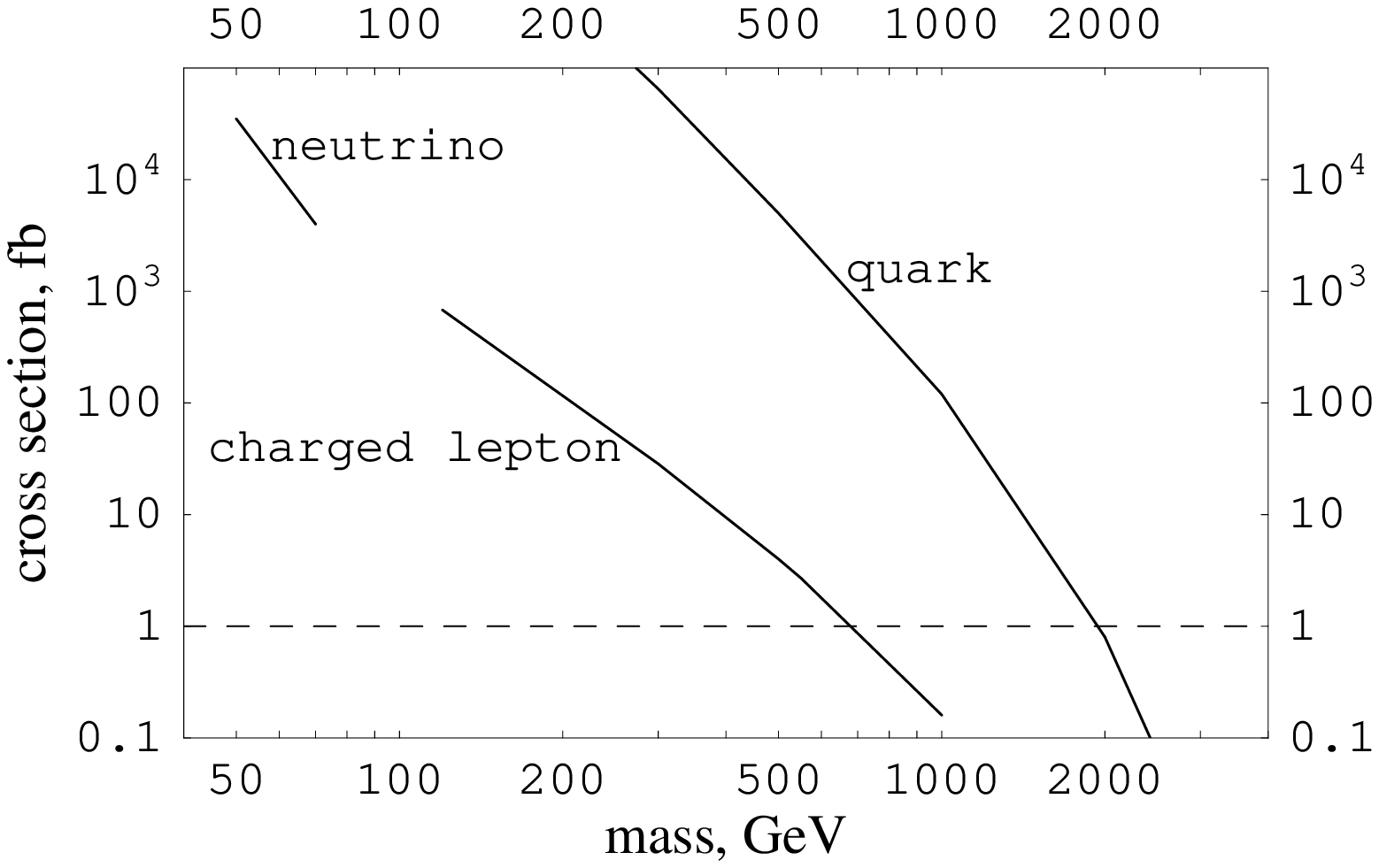}{Taken from \cite{lom} cross sections of
production of heavy stable quarks and leptons in experiment ATLAS.
Dash line shows the lower (conservative) level of LHC sensitivity
expected for 1st year of its operation.}

 \Acknow
{M.Kh. thanks LPSC (Grenoble, France) for hospitality and D.Rouable for help.}

\end{document}